\documentclass[aps,prd,10pt,twocolumn,showpacs,nofootinbib,preprintnumbers,amssymb]{revtex4-1}

\usepackage{graphicx, epsfig, amssymb} 
\usepackage{amsmath, amsfonts}
\usepackage{bm} 
\usepackage{pgf} 

\def\be{\begin{equation}}
\def\ee{\end{equation}}
\def\beq{\begin{eqnarray}}
\def\eeq{\end{eqnarray}}

\begin{document}

\centerline{}
\title{Gravitational perturbation of the BTZ black hole induced by test particles\\and weak cosmic censorship in AdS spacetime}

\author{
Jorge V. Rocha,$^{1}$
}
\email[Electronic address: ]{jorge.v.rocha@ist.utl.pt}
\author{
Vitor Cardoso$^{1,2}$
}
%
%
\affiliation{${^1}$ CENTRA, Departamento de F\'{\i}sica, 
Instituto Superior T\'ecnico, Universidade T\'ecnica de Lisboa,
Av.~Rovisco Pais 1, 1049 Lisboa, Portugal}
\affiliation{${^2}$ Department of Physics and Astronomy, The University of Mississippi, University, MS 38677, USA}

\date{\today}

\begin{abstract}
We analyze the gravitational perturbations induced by particles falling into a three dimensional, asymptotically AdS black hole geometry.
More specifically, we solve the linearized perturbation equations obtained from the geodesic motion of a ring-like distribution of test particles in the BTZ background.
This setup ensures that the $U(1)$ symmetry of the background is preserved.
The non-asymptotic flatness of the background raises difficulties in attributing the significance of energy and angular momentum to the conserved quantities of the test particles.
This issue is well known but, to the best of our knowledge, has never been addressed in the literature.
We confirm that the naive expressions for energy and angular momentum are the correct definitions.
Finally, we put an asymptotically AdS version of the weak cosmic censorship to a test:
by attempting to overspin the BTZ black hole with test particles it is found that the black hole cannot be spun-up past its extremal limit. 
\end{abstract}


\maketitle



\section{Introduction}

The cosmic censorship conjecture has been proposed by Penrose~\cite{Penrose:1969pc} in 1969 and despite its relatively long lifetime it remains one of the outstanding unresolved issues in classical general relativity.
The idea of a cosmic censor was put forward to prevent curvature singularities to be noticeable to distant observers.
Such an occurrence, termed naked singularity, would signal the break-down of predictability within the theory.
We shall be concerned only with the {\em weak} cosmic censorship conjecture (wCCC), which forbids the appearance of naked singularities as the endpoint of a generic gravitational collapse of physically acceptable matter starting off in a regular initial state.
A somewhat more precise formulation of the wCCC has been given in Ref.~\cite{Wald:1997wa}.

A notable attempt to destroy a black hole (BH) was envisaged by Wald in 1974~\cite{Wald:1974}.
The simplest such thought experiment consisted in throwing a point particle at a Kerr black hole, with large enough angular momentum to spin up the black hole.
The Kerr solution possesses an event horizon only when its angular momentum is bounded by its mass as $J\leq M^2$.
Therefore, if it were possible for the BH to absorb particles with sufficiently high angular momenta, then one might exceed the Kerr bound, thus creating a naked singularity.
However, the wCCC is unharmed by this test as the potentially dangerous particles are simply not captured by the black hole~\cite{Wald:1974}.

Following Wald's seminal work, many other attempts have been made to violate the wCCC~\cite{Hubeny:1998ga, Matsas:2007bj, Jacobson:2009kt, Lehner:2010pn} (see also~\cite{Hod:2008zza} and references therein).
Among the cases studied, those that succeeded in producing naked singularities either (i) neglected back-reaction or (ii) relied on the assumption of a high degree of symmetry throughout the gravitational evolution of the system.
Back-reaction effects have been argued to invalidate the analyses belonging to class~(i)~\cite{Hod:2008zza, Barausse:2010ka}, whereas the studies pertaining to class~(ii) do not correspond to generic situations.

Recently, Wald's analysis has been generalized to a multitude of higher dimensional spacetimes~\cite{BouhmadiLopez:2010vc}, more specifically the Myers-Perry~\cite{Myers:1986un} black holes and both neutral and dipole black rings~\cite{Emparan:2001wn, Emparan:2004wy} in five dimensions.
In all situations the findings were the same: none of the black holes considered were ever overspun.
A seemingly different problem, but in the same spirit, has been studied in~\cite{Gibbons:2008hq} where the authors attempted to speed up an infinite rigidly rotating dust cylinder by throwing in test particles.
Above a critical value of the rotation parameter the spacetime develops closed timelike curves and the impossibility of overspinning the system beyond this point was proven.

In this article we follow a program analogous to Wald's process in anti-de Sitter (AdS) backgrounds.
More precisely, we test an AdS version of the wCCC with the Ba\~nados-Teitelboim-Zanelli (BTZ) black hole~\cite{Banados:1992wn}.
In effect, the formulation of the wCCC presented in~\cite{Wald:1997wa} excludes the case of black holes in AdS simply because the notion of distant observers in non-asymptotically flat spacetimes is not well defined.
However, one might question whether this assumption is necessary to prevent the overspinning of black hole geometries.
Another motivation comes from AdS/CFT: black holes in AdS rotating above the extremal limit map to some state on the boundary theory rotating at a speed greater than light~\cite{Hawking:1998kw}.
Therefore, one should be distrustful about any physical process which would result in the overspinning of a black hole in AdS.

We thus propose to investigate the problem of spinning up a rotating black hole in AdS with test particles.
The setting for our analysis shares some similarities with a previous study~\cite{Wang:1996hp}.
However, a crucial issue was left unanswered:
how does one define {\it energy} and {\it angular momentum} of the test particles?
A rotating stationary spacetime specified by a metric tensor $g_{\mu\nu}$ possesses both timelike and rotational Killing vectors, from which we can build two conserved quantities,
\beq
E &\equiv& -g_{\mu\nu} (\partial/\partial t)^\mu \dot z^\nu = -\frac{dT}{d\tau}\left[ g_{tt}+g_{t\phi}\frac{d\Phi}{dt} \right]\,, \label{eq:EandL1}\\
L &\equiv& g_{\mu\nu} (\partial/\partial \phi)^\mu \dot z^\nu = \frac{dT}{d\tau}\left[ g_{t\phi}+g_{\phi\phi}\frac{d\Phi}{dt} \right]\,,
\label{eq:EandL2}
\eeq
where $z^\mu(\tau)=(T(\tau),R(\tau),\Phi(\tau))$ stands for the particle coordinates and the dot indicates derivation with respect to proper time $\tau$.
In an asymptotically flat spacetime these would correspond to the energy and angular momentum of the test particles.
However, in asymptotically AdS geometries this identification cannot be immediately made.
Of course, this issue has been known for many decades but, to the best of our knowledge, has never been addressed in the literature.
One of the main points of this article is to show that $E$ and $L$ given above are indeed the correct expressions.

The strategy we will follow to identify the conserved quantities is to consider linear perturbations induced by test particles.
This was first done by Zerilli~\cite{Zerilli:1971wd} for the Schwarzschild solution, building on the seminal work by Regge and Wheeler ~\cite{Regge:1957td}.
The presence of such test particles in the geometry will produce a variation (at the linear level) of the mass and angular momentum of the black hole which defines the background.
We then read off $E$ and $L$ from these variations.
References~\cite{Zerilli:1971wd, Regge:1957td} rely heavily on the high degree of symmetry of the Schwarzschild black hole.
Of course, once one considers rotating black holes the amount of symmetry gets reduced and the problem becomes technically less tractable.
The only exception occurs in three spacetime dimensions and this is the reason why we consider the BTZ black hole.
Furthermore, to preserve the rotational symmetry of the background we will perturb the geometry, not with a point particle but with a circular homogeneous distribution of test particles.
This amounts to keeping only the zero-modes in our analysis, i.e., we shall consider metric perturbations independent of the angular coordinate $\phi$.

The plan of the paper is as follows.
In Sec.~\ref{sec:Prelim} the general form of the linearized perturbation equations for asymptotically AdS spacetimes is given and the stress-energy tensor describing the infall of a ring-like distribution of test particles in the BTZ geometry is obtained.
The gravitational perturbation equations are then solved in Sec.~\ref{sec:PertBTZ}.
The result of this calculation allows us to identify the energy and angular momentum of the test particles.
Finally, in Sec.~\ref{sec:SpinUp} it is shown that the BTZ black hole cannot be overspun by throwing in test particles.
The issues concerning gauge transformations are relegated to the Appendix.

\section{Preliminaries
\label{sec:Prelim}}

\subsection{Gravitational perturbations of asymptotically AdS solutions}

Consider a small perturbation $h_{\mu\nu}$ of a background metric $g_{\mu\nu}$,
\be
\widetilde{g}_{\mu\nu}=g_{\mu\nu}+h_{\mu\nu}\,.
\ee
We assume the background metric is a solution of the sourceless cosmological Einstein equations in $D$ spacetime dimensions,
\be
G_{\mu\nu}+\frac{D-2}{2}\Lambda g_{\mu\nu}=0 \,,
\ee
where $G_{\mu\nu}=R_{\mu\nu}-\frac{1}{2}g_{\mu\nu}R$ denotes the Einstein tensor and $\Lambda \equiv -(D-1)/\ell^2$ is the (negative) cosmological constant.
Then, the linearized perturbation equations read as follows~\cite{Ortin:2004ms}:
\begin{align}
2\delta G_{\mu\nu} &\equiv -\nabla^2h_{\mu\nu} +2\nabla^\lambda \nabla_{(\mu}h_{\nu)\lambda} -\nabla_\mu \nabla_\nu h  \nonumber\\
	& +g_{\mu\nu} \left(\nabla^2 h - \nabla^\alpha \nabla^\beta  h_{\alpha\beta} \right) -h_{\mu\nu}R  \nonumber\\
  & +g_{\mu\nu} h_{\alpha\beta}R^{\alpha\beta} + (D-2)\Lambda h_{\mu\nu} = 16\pi G_N T_{\mu\nu}\,,
\label{eq:linEinstein}
\end{align}
where the covariant derivatives are taken with respect to the background metric and $h\equiv g^{\mu\nu}h_{\mu\nu}$ denotes the trace of the metric perturbation.
$T_{\mu\nu}$ is the stress-energy tensor of the test particles that will drive the perturbation.
By virtue of the Bianchi identity, the stress-energy tensor must be divergenceless to ensure consistency of Eqs.~\eqref{eq:linEinstein}.
This occurs if and only if the source particles follow geodesics.

In the remainder of the manuscript we shall particularize to $D=3$.

\subsection{The BTZ black hole
\label{sec:BTZ}}

The BTZ solution \cite{Banados:1992wn} describes an asymptotically anti-de Sitter rotating black hole in $D=3$ spacetime dimensions. 
The spacetime is defined by the line element
\be
ds^2=-N^2dt^2+N^{-2}dr^2+r^2\left(N^{\phi}dt+d\phi\right)^2\,,
\label{eq:BTZmetric}
\ee
with lapse (squared) and shift
\be
N^2=-M+\frac{r^2}{\ell^2}+\frac{J^2}{4r^2}\,, \qquad\;  N^{\phi}=-\frac{J}{2r^2}\,.
\ee
Here, $M$ and $J$ are the total mass and angular momentum of the spacetime, and $\ell\;(>0)$ is the anti-de Sitter radius.
The spacetime has a Cauchy and an event horizon at
\be
\frac{2r_{\pm}^2}{\ell^2}=M\left(1\pm\sqrt{1-\frac{J^2}{M^2\ell^2}}\right)\,.
\ee
Therefore, the existence of an event horizon imposes a bound on the angular momentum $J$, such that
\be
|J|\leq M\ell\,.
\label{eq:boundBTZ}
\ee
%

\subsection{The stress-energy tensor}

For a single test particle of rest-mass $m_0$ the stress-energy tensor is given by
\beq
T^{\mu\nu}_{(1p)} &=& m_0\int \delta^{(3)}(x-z(\tau))\frac{dz^{\mu}}{d\tau}\frac{dz^{\nu}}{d\tau}d\tau\,, \nonumber\\
                  &=& m_0\frac{dT}{d\tau}\frac{dz^{\mu}}{dt}\frac{dz^{\nu}}{dt}\frac{\delta(r-R(t))}{r}\delta(\phi-\Phi(t))\,,
\eeq
where $\tau$ denotes proper time along the world line $z^{\mu}(\tau)$.

As mentioned before, we will perturb the BTZ spacetime with test particles homogeneously distributed along a ``ring''.
The stress-energy tensor is then
\be
T^{\mu\nu} = m_0\frac{dT}{d\tau}\frac{dz^{\mu}}{dt}\frac{dz^{\nu}}{dt}\frac{\delta(r-R(t))}{r}\,,
\ee
and the quantity $m_0$ may be viewed as the mass density of the ring.

The components of the stress-energy tensor may be expressed as follows:
\beq
T_{tt} &=& -m_0 E \left[ g_{tt}+g_{t\phi}\frac{d\Phi}{dt} \right] \frac{\delta(r-R(t))}{r}\,, \nonumber\\
T_{tr} &=& -m_0 E g_{rr} \frac{dR}{dt} \frac{\delta(r-R(t))}{r}\,, \nonumber\\
T_{t\phi} &=& -m_0 E \left[ g_{t\phi}+g_{\phi\phi}\frac{d\Phi}{dt} \right] \frac{\delta(r-R(t))}{r} \nonumber\\
          &=& m_0 L \left[ g_{tt}+g_{t\phi}\frac{d\Phi}{dt} \right] \frac{\delta(r-R(t))}{r}\,, \nonumber\\
T_{rr} &=& m_0 \frac{dT}{d\tau} g_{rr}^2 \left(\frac{dR}{dt}\right)^2 \frac{\delta(r-R(t))}{r}\,, \nonumber\\
T_{r\phi} &=& m_0 L g_{rr} \frac{dR}{dt} \frac{\delta(r-R(t))}{r}\,, \nonumber\\
T_{\phi\phi} &=& m_0 L \left[ g_{t\phi}+g_{\phi\phi}\frac{d\Phi}{dt} \right] \frac{\delta(r-R(t))}{r}\,.
\label{eq:stresstensor}
\eeq
%

\section{Linearized perturbation equations for BTZ
\label{sec:PertBTZ}}

As shown in the Appendix, the existing gauge freedom allows us to eliminate several components of the metric perturbation so that we can choose it to take the following form:
\be
h_{\mu \nu}= \left[
 \begin{array}{ccc}
  A(t,r) & 0          & W(t,r)\\
  0      & B(t,r)/N^4 & 0     \\
  W(t,r) & 0          & 0
\end{array}\right] \,.
\label{eq:gaugefix}
\ee
With this choice of gauge, the trace of the metric perturbation becomes
\be
h = H(t,r) \equiv \frac{-A(t,r)+B(t,r)+2N^\phi W(t,r)}{N^2}\,.
\label{eq:trace}
\ee

Given the stress-energy tensor~\eqref{eq:stresstensor}, the $\{tr\}$ and $\{r\phi\}$ components of the linearized cosmological Einstein Eq.~\eqref{eq:linEinstein} become
\begin{align}
\frac{1}{rN^2}\partial_t & \left\{ -r^2(N^\phi)^2 H + B - r\partial_r (N^\phi W) \right\} \nonumber\\
	 & \qquad\qquad\qquad = -\kappa\, m_0 E \frac{dR}{dt} \frac{\delta(r-R(t))}{rN^2}\,,\label{eq:Ein12}\\
\frac{1}{rN^2}\partial_t & \left\{ -r^2 N^\phi H - r^3\partial_r \left(\frac{W}{r^2}\right) \right\} \nonumber\\
	 & \qquad\qquad\qquad = \kappa\, m_0 L \frac{dR}{dt} \frac{\delta(r-R(t))}{rN^2}\,,\label{eq:Ein23}
\end{align}
where we have defined $\kappa \equiv 16\pi G_N$ for convenience.
Performing the time integration one obtains
\begin{align}
&-r^2(N^\phi)^2 H + B - r\partial_r (N^\phi W) = \kappa\, m_0 E\, \Theta(r-R(t))\,,\label{eq:PartialSol1}\\
&-r^2 N^\phi H - r^3\partial_r \left(\frac{W}{r^2}\right) = -\kappa\, m_0 L\, \Theta(r-R(t))\,,\label{eq:PartialSol2}
\end{align}
where $\Theta$ represents the Heaviside step function.
Multiplying~\eqref{eq:PartialSol2} by $N^\phi$ and subtracting from~\eqref{eq:PartialSol1} immediately yields
\be
B(t,r) = \kappa\, m_0 (E+N^\phi L)\, \Theta(r-R(t))\,.
\label{eq:SolB}
\ee

Now consider the following linear combinations of the components of Eq.~\eqref{eq:linEinstein}:
$\{tt\}-N^\phi\{t\phi\}$ and $\{t\phi\}-N^\phi\{\phi\phi\}$.
The resulting equations are
\begin{align}
\frac{N^2}{r}\partial_r & \left\{ -r^2(N^\phi)^2 H + B - r\partial_r (N^\phi W) \right\} \nonumber\\
	 & \qquad\qquad\qquad = \kappa\, m_0 E \frac{N^2}{r} \delta(r-R(t))\,,\label{eq:Ein13and33}\\
\frac{N^2}{r}\partial_r & \left\{ -r^2 N^\phi H - r^3\partial_r \left(\frac{W}{r^2}\right) \right\} \nonumber\\
	 & \qquad\qquad\qquad = -\kappa\, m_0 L \frac{N^2}{r} \delta(r-R(t))\,,\label{eq:Ein11and13}
\end{align}
It is easy to see that after a radial integration these equations yield precisely the same solutions~\eqref{eq:PartialSol1} and~\eqref{eq:PartialSol2} as above.

There remain two equations out of the linearized Einstein equations to be analyzed, which we may choose to be the $\{rr\}$ and $\{\phi\phi\}$ components.
However, the six equations are not all independent: setting the divergence of both sides of Eq.~\eqref{eq:linEinstein} to zero allows one to express the $\{\phi\phi\}$ component in terms of the other linear combinations we are considering (and derivatives thereof).
This leaves us with just one more independent equation to consider, namely the $\{rr\}$ component:
\be
-\frac{2}{\ell^2N^2}H - \frac{1}{rN^2}\partial_r A 
= \kappa\, m_0 \frac{dT}{d\tau} \left(\frac{dR}{dt}\right)^2 \frac{\delta(r-R(t))}{rN^4}\,.
\ee
Multiplying the above by $rN^2$ and subtracting Eq.~\eqref{eq:PartialSol2} times $2N^\phi/r$ yields
\begin{align}
&N^2\partial_r H - \partial_r B \nonumber\\
& =\kappa\, m_0 \left[ \frac{dT}{d\tau}\left(\frac{dR}{dt}\right)^2\frac{\delta(r-R(t))}{N^2} + \frac{2N^\phi}{r}L\Theta(r-R(t)) \right]\,.
\end{align}
By replacing the solution for $B(t,r)$ previously found in~\eqref{eq:SolB} it can be verified that $\partial_r H \propto \delta(r-R(t))$.
Therefore, an integration in $r$ results in $H(t,r)$ being a function of time only, for $r>R(t)$.
More specifically, one obtains
\be
H(t,r) = \kappa\, m_0 F(t)\Theta(r-R(t))\,,
\label{eq:SolH}
\ee
where we have defined
\be
F(t) \equiv \left[2\frac{E+N^{\phi}L}{N^2}-\frac{1}{E+N^{\phi}L}\left(\epsilon+\frac{L^2}{r^2}\right)\right]_{r=R(t)}\,.
\ee
Here, the quantity $\epsilon$ arises from a term proportional to $\dot{R}^2$; see Eq.~\eqref{eq:radialequation} below.

We can now insert~\eqref{eq:SolH} back into~\eqref{eq:PartialSol2} to find that
\begin{align}
W(t,r) &= -\kappa\, m_0\frac{L}{2}\Theta(r-R(t)) +\kappa\, m_0\Theta(r-R(t))\nonumber\\
 \times & \left[ -\frac{J F(t)}{4} +\frac{r^2}{R(t)^2}\left(\frac{L}{2}+\frac{J F(t)}{4}\right) \right]\,.
\label{eq:SolW}
\end{align}

Now, putting together Eqs.~\eqref{eq:trace}, \eqref{eq:SolB}, \eqref{eq:SolH} and~\eqref{eq:SolW} immediately gives
\begin{align}
A(t,r) &= \kappa\, m_0 E \Theta(r-R(t)) +\kappa\, m_0\Theta(r-R(t))\nonumber\\
   &\!\!\!\!\!\!\!\! \times \left[ \left(M-\frac{r^2}{\ell^2}\right)F(t)-\frac{J}{R(t)^2}\left(\frac{L}{2}+\frac{J F(t)}{4}\right)\right]\,.
\label{eq:SolA}
\end{align}

Finally, evoking the results from the Appendix, we can eliminate the terms inside square brackets in Eqs.~\eqref{eq:SolW} and~\eqref{eq:SolA} simultaneously by making a residual gauge transformation.
This same gauge transformation sets $H(t,r)=0$, in agreement with Eq.~\eqref{eq:trace}.

Hence, the solution for the metric perturbation can be cast into the form~\eqref{eq:gaugefix} with the non-vanishing components given by
\beq
A(t,r) &=& \kappa\, m_0 E \Theta(r-R(t))\,, \nonumber\\
W(t,r) &=& -\kappa\, m_0\frac{L}{2}\Theta(r-R(t))\,, \nonumber\\
B(t,r) &=& \kappa\, m_0 (E+N^\phi L)\, \Theta(r-R(t))\,.
\eeq
The result we have obtained for the linearized perturbation agrees with our expectations:
referring to Eq.~\eqref{eq:BTZmetric} we conclude that
{\em the perturbation imparted by the circular distribution of test particles vanishes inside the ring and 
corresponds simply to shifting the charges of the BTZ black hole in the outer region},
\be
M\rightarrow M+\delta M\,, \qquad\quad J\rightarrow J+\delta J\,.
\ee
The precise values we have found for the shifts,
\be
\delta M = 16\pi G_N m_0 E\,, \qquad\quad \delta J = 16\pi G_N m_0 L\,,
\ee
justify the naive identification of $m_0E$ and $m_0L$, as expressed in Eqs.~\eqref{eq:EandL1} and~\eqref{eq:EandL2}, with the energy and angular momentum of the test particles.
As usual, for massive particles $E$ and $L$ may be interpreted as energy and angular momentum per unit mass.

\section{Spinning-up the BTZ black hole by throwing point particles
\label{sec:SpinUp}}

From the above, we now know how to relate physical mass and angular momentum of an infalling particle to its conserved quantities.
Let us try to spin-up a BTZ black hole with mass $M_0$ and angular momentum $J_0$.
For that, we throw in a particle of mass $m_0$ with angular momentum $\delta J=m_0L$ and energy $\delta M=m_0E$.
According to the extremality bound~\eqref{eq:boundBTZ} the dimensionless spin of the BH must satisfy $|j_0|\equiv J_0/(M_0\ell) \leq 1$.

To be able to clearly define the problem, we will ask for well-defined initial and final states.
Thus, initially we consider a particle on an unbound geodesic and sufficiently far away from the black hole.
This is equivalent to really throwing in a particle from ``infinity''~\footnote{It is not clear to us how to have a well {\it understood} initial state when the point particle is on a bound geodesic, i.e., how to define the ``black hole'' and ``the particle'' in such a situation.}.

Now, the only particles on unbound geodesics must be null~\cite{Cruz:1994ir}. In fact, it is not hard to show that
the geodesic equations yield
\beq
\dot{\Phi}  &=& \frac{2E J\ell^2-4\ell^2LM+4LR^2}{4R^2\ell^2N^2}\,,\\
\dot{T}     &=& \frac{2ER^2- JL}{2R^2N^2}\,,\\
\dot{R}^2   &=& -\epsilon N^2+E^2-L^2/\ell^2+R^{-2}(L^2M- JEL)\,,\label{eq:radialequation} 
\eeq
where dots denote proper time derivatives and $\epsilon=1,0$ for timelike or null geodesics, respectively.
In these equations the lapse should be viewed as a function of $R$ instead of $r$.

The radial Eq.~\eqref{eq:radialequation} can be integrated directly, and it is easy to see that massive particles in the BTZ geometry are {\it always} on bound orbits and are always captured by the black hole~\cite{Farina:1993xw,Cruz:1994ir}.
An inspection of the radial Eq.~\eqref{eq:radialequation} shows that for the particle to be on an unbound orbit, it must be null and satisfy the requirement~\cite{Cruz:1994ir}
\be
q\equiv \frac{L}{E\ell} \leq 1\,.
\label{eq:limitbtz}
\ee
Now, upon absorption of a particle, the dimensionless spin of a BTZ black hole becomes
\be
j\equiv \frac{J}{M\ell}=\frac{J_0+m_0\,L}{\ell(M_0+m_0\,E)}=\frac{j_0+qm_0E/M_0}{1+m_0E/M_0}\,.
\ee
Clearly, $j_0,q\leq 1$ implies $j\leq 1$, with the equality being satisfied in the extremal limit if and only if $L=E\ell$.
Thus, a BTZ black hole cannot be spun-up past extremality.
Unlike in previous studies~\cite{Jacobson:2009kt,BouhmadiLopez:2010vc}, the above derivation is quite general and does not rely explicitly on an expansion around $m_0E=0$, i.e., the point particle limit
(though it is assumed that the particle follows a geodesic, as imposed by the requirement of consistency of the linearized perturbations).

\subsection{The massive case}
Our argument above for considering only null particles relies on the fact that having a well defined initial state requires the particle to be sent from infinity.
However, one could argue that in the limit that $m_0\to 0$, one should be allowed to describe a system in terms of a black hole plus a point particle,
since the back-reaction on the geometry tends to zero in this limit.
We now show that in this regime massive particles cannot over-spin the black hole.
The condition that the geodesic is future-directed, $\dot{T}\geq0$, constrains the angular momentum to satisfy (see Eqs.~\eqref{eq:radialequation}) $ER^2-\frac{1}{2}JL\geq0$ outside the event horizon.
At the horizon, this condition implies that 
\be
\frac{L}{E}\leq \frac{2r_+^2}{J}\,.
\label{eq:limitbtz2}
\ee
For an extreme black hole this yiels $L/E\leq \ell$.
Upon absorption of a particle, the dimensionless spin of the BTZ black hole is 
\be
j = j_0-\frac{m_0E}{M_0}\left(j_0-\frac{L}{E\ell}\right)\,.
\ee
Notice an important difference with respect to the null particle calculation: we are now Taylor-expanding around $j_0$, because we are only allowed to work in the regime $m_0\to 0$. For $j_0=1$, taking relation (\ref{eq:limitbtz2}) into account yields $j\leq 1$.
Thus, a BTZ black hole cannot be spun-up past extremality.

\section{Conclusions}

We have shown how to read off the physical energy and angular momentum of point particles falling into a BTZ black hole.
The results of our calculation and generalizations thereof are interesting for a number of processes taking place in the vicinities of asymptotically anti-de Sitter black holes.
We have studied a particular attempt at violating the weak version of the Cosmic Censorship Conjecture, an attempt at overspinning the BTZ black hole by throwing point particles into it.
We argued that for well-defined initial conditions, BTZ black holes {\it cannot} be spun up past extremality, thus verifying the wCCC in the non-trivial case of asymptotic AdS spacetimes. 

The technical problem we have tackled, namely the analysis of gravitational perturbations of rotating black holes, naturally led us to consider black holes in $(2+1)$ dimensions.
In this lower-dimensional case we can take advantage of the existing amount of symmetry of the problem to completely solve for the perturbation.
Nevertheless, there is a bonus: in three spacetime dimensions there are no gravitational-wave degrees of freedom and so back-reaction effects are suppressed relatively to the four-dimensional, asymptotically flat case~\cite{Jacobson:2009kt,Barausse:2010ka}.
Therefore, gravitational radiation could not have been evoked to prevent overspinning (if it were possible) and our results should be the end of the story.
It would be highly desirable to extend this study to generic asymptotically AdS$_D$ geometries, even though the path followed here may not have a simple generalization.

\section*{Acknowledgements}

We thank Jos\'e Nat\'ario for bringing Ref.~\cite{Gibbons:2008hq} to our attention.
JVR is supported by {\it Funda\c{c}\~ao para a Ci\^encia e Tecnologia} (FCT)-Portugal through Contract No. SFRH/BPD/47332/2008.
This work was supported by the {\it DyBHo--256667} ERC Starting Grant and by FCT - Portugal through PTDC projects FIS/098025/2008, FIS/098032/2008, CTE-AST/098034/2008.

\section*{Appendix: Gauge transformations
\label{sec:gauge}}

Under rotations around the origin, the six components of the perturbation tensor transform like three scalars, namely $h_{tt}$, $h_{tr}$ and $h_{rr}$,
one vector $(h_{t\phi},h_{r\phi})$ and one second-order tensor $h_{\phi\phi}$.
The scalars can be decomposed into their Fourier modes (labeled by $n$).
To preserve the $U(1)$ symmetry of the background geometry we will consider a perturbation by a circular homogeneous distribution of test particles.
Thus, only the $\phi$-zero-modes survive.

For the ``vector'' components there are only two quantities in the problem with the right transformation properties:
a constant vector times $\partial_\phi e^{i n \phi}$; and $(g_{t\phi},g_{r\phi})$.
Since by construction we pick only the $n=0$ modes, the first candidate vanishes.
From~\eqref{eq:BTZmetric} we then have $h_{r\phi}=0$ automatically.

Under an infinitesimal coordinate transformation, $x^\mu \rightarrow x^\mu-\frac{1}{2}\xi^\mu$, the metric perturbation transforms as
\be
h_{\mu\nu} \longrightarrow h_{\mu\nu}^{\rm new} = h_{\mu\nu} + \nabla_{(\mu}\xi_{\nu)}\,.
\ee
In order not to spoil the rotational invariance we wish to consider only gauge transformations that do not depend on the angle $\phi$.
Thus,
\beq
h_{tr}^{\rm new}        &=&  h_{tr} + \frac{1}{2}\partial_t\xi_r + \frac{1}{2}\partial_r\xi_t - \frac{r}{\ell^2 N^2} \left( \xi_t - N^\phi \xi_\phi \right) \,, \nonumber\\
h_{r\phi}^{\rm new}     &=&  h_{r\phi} + \frac{r^2}{2}\partial_r\left(\frac{\xi_\phi}{r^2}\right) - \frac{r N^\phi}{N^2} \left( \xi_t - N^\phi \xi_\phi \right) \,, \nonumber\\
h_{\phi\phi}^{\rm new}  &=&  h_{\phi\phi} + r N^2 \xi_r\,.
\eeq

First we make a gauge transformation of the form $\xi_\mu^{(1)} = (0,\xi_r,0)$.
With the choice $\xi_r(t,r) =-h_{\phi\phi}/r N^2$ we set $h_{\phi\phi}^{\rm new}=0$.
This transformation preserves $h_{r\phi}=0$.

Next, consider a coordinate transformation implemented by $\xi_\mu^{(2)} = (\xi_t,0,\xi_\phi)$.
This preserves $h_{\phi\phi}=0$.
To keep $h_{r\phi}=0$ as well, we need
\be
\xi_t = N^\phi \xi_\phi + \frac{r N^2}{2 N^\phi}\partial_r\left(\frac{\xi_\phi}{r^2}\right) \,.
\label{eq:gauge23}
\ee
To eliminate the component $h_{tr}$ we must have
\be
h_{tr} + \frac{1}{2}\partial_r\xi_t - \frac{r}{\ell^2 N^2} \left( \xi_t - N^\phi \xi_\phi \right) = 0 \,.
\ee
Replacing $\xi_t$ by~\eqref{eq:gauge23} we obtain
\be
\partial_r \left( r^3 \partial_r \left(\frac{\xi_\phi}{r^2}\right) \right) = \frac{2J}{N^2}h_{tr} \,,
\ee
for which the solution is
\be
\xi_\phi = r^2 \int_r^\infty r^{-3} \left( \int_r^\infty \frac{2J}{N^2}h_{tr}dr \right)dr \,.
\label{eq:gauge12}
\ee
Thus, the gauge transformation described by $\xi^{(2)}$ with $\xi_\phi$ and $\xi_t$ given respectively by~\eqref{eq:gauge12} and~\eqref{eq:gauge23} effectively sets $h_{tr}^{\rm new}=h_{r\phi}^{\rm new}=0$.

In accordance with the above considerations, we choose to work in a gauge such that the metric perturbation takes the form~\eqref{eq:gaugefix}.

At this point note that there is a further {\em residual} gauge freedom: one can still make a coordinate transformation of the form 
$\xi_\mu^{(3)} = (\xi_t^{\rm res},0,\xi_\phi^{\rm res})$ with
\beq
\xi_\phi^{\rm res} &=&  f_1(t)+r^2f_2(t)\,, \nonumber\\
\xi_t^{\rm res}    &=&  -\frac{2}{J}\left(M-\frac{r^2}{\ell^2}\right)f_1(t)-\frac{J}{2}f_2(t)\,.
\eeq
This leaves the components $h_{tr}$, $h_{rr}$, $h_{r\phi}$ and $h_{\phi\phi}$ unaltered, while the remaining components and the trace change as
\beq
h_{tt}^{\rm new}     &=&  h_{tt} -\frac{2}{J}\left(M-\frac{r^2}{\ell^2}\right)f_1'(t)-\frac{J}{2}f_2'(t)\,, \nonumber\\
h_{t\phi}^{\rm new}  &=&  h_{t\phi} +\frac{1}{2}f_1'(t) +\frac{r^2}{2}f_2'(t)\,, \nonumber\\
h^{\rm new}          &=&  h -\frac{2}{J}f_1'(t)\,.
\eeq
Therefore, we can use the residual gauge transformations to eliminate a function of time only from the trace of the metric perturbation 
and to eliminate terms of the form $F_1(t)+r^2 F_2(t)$ from $h_{tt}$ and $h_{t\phi}$.
In Sec.~\ref{sec:PertBTZ} we exploit this residual gauge freedom when solving the linearized perturbation equations.


\end{document}